# US News and Social Media Framing around Vaping

Keyu Chen[1], Marzieh Babaeianjelodar[1], Yiwen Shi[2], Rohan Aanegola[1], Lam Yin Cheung[3], Preslav Ivanov Nakov[4], Shweta Yadav[5], Angus Bancroft[6], Ashiqur R. Khudabukhsh[7], Munmun De Choudhury[8], Frederick L. Altice[1], and Navin Kumar[1,9]

1. Yale University School of Medicine, New Haven, CT 06510, USA
2. Yale University, New Haven, CT 06510, USA
3. FTI Consulting, Inc
4. Qatar Computing Research Institute
5. University of Illinois Chicago
6. The University of Edinburgh, College of Arts
7. Rochester Institute of Technology
8. Georgia Institute of Technology
9. National University of Singapore

**Abstract.** In this paper, we investigate how vaping is framed differently (2008-2021) between US news and social media. We analyze 15,711 news articles and 1,231,379 Facebook posts about vaping to study the differences in framing between media varieties. We use word embeddings to provide two-dimensional visualizations of the semantic changes around vaping for news and for social media. We detail that news media framing of vaping shifted over time in line with emergent regulatory trends, such as; flavored vaping bans, with little discussion around vaping as a smoking cessation tool. We found that social media discussions were far more varied, with transitions toward vaping both as a public health harm and as a smoking cessation tool. Our cloze test, dynamic topic model, and question answering showed similar patterns, where social media, but not news media, characterizes vaping as combustible cigarette substitute. We use n-grams to detail that social media data first centered on vaping as a smoking cessation tool, and in 2019 moved toward narratives around vaping regulation, similar to news media frames. Overall, social media tracks the evolution of vaping as a social practice, while news media reflects more risk based concerns. A strength of our work is how the different techniques we have applied validate each other. Stakeholders may utilize our findings to intervene around the framing of vaping, and may design communications campaigns that improve the way society sees vaping, thus possibly aiding smoking cessation; and reducing youth vaping.

# 1 Introduction

The recent introduction of alternative forms of nicotine products into the marketplace (e.g., e-cigarettes/vapes, heated tobacco products, and smokeless tobacco) has led to a more complex informational environment [17]. The scientific consensus is that vape aerosol contains fewer numbers and lower levels of toxicants than smoke from combustible tobacco cigarettes [26]. Among youth in the USA, the adolescent nicotine vaping use increased from 2017 to 2019 but then started declining in 2020, which includes a decline in daily vaping as well [21]. Among adults, a Cochrane review found that nicotine vapes probably do help people to stop smoking for at least six months, and working better than nicotine replacement therapy and nicotine-free e-cigarettes [9]. Given that vaping is represented in the public health environment both as a smoking cessation tool and as a harm to youth health, it is highly controversial and polarizing. Early divisions among public health experts led to polarized coverage in the media, confused messages to the public, and inconsistent policymaking between jurisdictions [30]. For many authorities in the United States, the potential health harms of vaping and youth-vaping concerns are overriding considerations of vaping as a smoking cessation tool [32]. The coverage of controversial topics, such as vaping, is often different in social media versus in news media. For example, social media can be more hyperbolic compared to news media [13], or can be used primarily to share information rather than to report news events. News media can shape public perceptions and policy about vaping [31], particularly in the context of major vaping events, such as the August 2019 outbreak of vaping product useassociated lung injury (EVALI). In late 2019, the Centers for Disease Control and Prevention (CDC) began to investigate a steep rise in hospitalizations linked to vaping product use. The condition came to be called EVALI. Researchers identified vitamin E acetate, a chemical added to some THC-containing vaping products, as the main cause of the illness. However, news reports did not always differentiate between THC devices and standard nicotine-based vapes [13], perhaps disproportionately characterizing vaping harms. Such vaping-related news may have triggered national and state-level policy responses, and may have influenced public perceptions (including misperceptions) regarding the harms of vaping [14]. Social media such as Facebook, Twitter, and YouTube have recently become important platforms for health surveillance and social intelligence [4], providing new insights on vaping to help inform future research, regulations, surveillance, and enforcement efforts. For example, vape tricks, e.g., blowing large vapor clouds or shapes like rings, are popular content on vaping-related social media [16], which can provide insight into risky tobacco use practices. Comparisons between how social media and news media frame vaping are key to managing vaping-related health outcomes. By identifying the differences in such media framing, we can design targeted policy tools specific to each type of medium. For example, messages around curbing youth vaping may be effective on social media, given the proliferation of youth-related vaping content. However, ads on news media designed to mitigate youth vaping may be ineffective as these are not platforms frequented by youth and are perhaps more suited for promoting vaping as a cessation tool for adults who already smoke and wish to quit smoking. Thus, capturing the differences in how vaping is framed on social media vs news media are key to designing targeted policy tools to simultaneously reduce youth vaping and to improve smoking cessation rates, thereby minimizing vape harms and maximizing benefits.

Most research around news media and vaping centered on responses to the 2020 outbreak of vaping-related lung injury (EVALI) [12]. Research regarding vaping on social media generally falls under two domains: machine learning to assist smoking cessation, and content analysis of vaping on social media [6]. While past work provides detail around how vaping is framed on news and social media, there is limited analysis pre-EVALI, and minimal comparisons between news and social media. Thus, we propose a study comparing US social media to US news media framing of vaping. We use Media Cloud to obtain news media articles, and CrowdTangle for Facebook posts, cognizant that these do not provide a representative view of public discourse. Our main research question (RQ) is thus as follows: What are the broad differences between news media (Media Cloud data) and Facebook regarding vaping? Findings suggest that social media tracks the evolution of vaping as a social practice, while news media reflects more risk based concerns.

## 2    Related Work

**Vaping Discussion on News Media** Most research around news media and vaping centered on responses to the 2020 outbreak of vaping-related lung injury (EVALI) [22]. Generally, research articles on vaping in news media tend to frame vaping as a risk [29], possibly reducing youth health outcomes. A study found that the outbreak of EVALI was related to increased news coverage about the dangers of vaping and internet searches for vaping cessation [18]. Similar work gathered all articles published in US news sources in 2019 to find that vapingrelated news coverage was high, driven largely by EVALI [14]. Other work indicated that news events were associated with US vape sales, measured by retail sales data [12]. It is important to better understand how vaping is framed on news media, as such media can shape public perceptions, sales and policy. Our work provides examples of how news media frames vaping. Our findings can facilitate policy interventions that can be harnessed by stakeholders to improve smoking cessation outcomes.

**Vaping Discussion on Social Media** Research regarding vaping on social media generally falls under two domains: machine learning to assist smoking cessation [25], and; content analysis of vaping on social media [6]. Generally, research articles on vaping in social media tend to frame vaping as an increasing health harm, and also as a major contributor to the youth vaping crisis. Recent work used unsupervised machine-learning to categorize vaping images on Instagram [15], and BERT to model the sentiment of vaping tweets [20]. It is clear that social media are essential to surveillance and enforcement around vaping. Thus, exploring how vaping is framed on social media is central to maximizing health benefits from vapes and for minimizing youth use. In this paper, we provide examples of social media frame vaping. Our results can aid media interventions aiming to limit youth vaping use.

## 3   METHODS

**Data** To capture news media around vaping, we used the open-source media analysis platform Media Cloud (mediacloud.org) to analyze 15,711 media articles between 2008 and 2021 from 271 US media sources. For locating articles related to vaping, we used queries based on a related systematic review [1]: electroniccigarette, electronic cigarette, electronic cig, e-cig, ecig, e cig, e-cigarette, ecigarette, e cigarette, e cigar, e-juice, ejuice, ejuice, e-liquid, eliquid, e liquid, esmoke, esmoke, e smoke, vape, vaper, vaping, vape-juice, vape-liquid, vapor, vaporizer, boxmod, cloud chaser, cloudchaser, smoke assist, ehookah, e-hookah, e hookah, smoke pod, e-tank, electronic nicotine delivery system. We analyzed the articles from 24811 URLs (9100 URLs were broken links) from Media cloud. For social media, we used the CrowdTangle service (crowdtangle.com) to obtain 1,231,379 Facebook posts from 2009 to 2021, with the same set of keywords. The CrowdTangle database includes: 7M+ Facebook pages, groups, and verified profiles. This includes all public Facebook pages with more than 50K likes, all public Facebook groups with 95k+ members, all US-based public groups with 2k+ members, and all verified profiles. The Facebook data we have provided is not solely US-based and contains posts worldwide. Two reviewers independently examined 100 articles to confirm the salience with our research question. The reviewers then discussed their findings and highlighted items deemed relevant across both lists, determining that 95% were relevant. As Facebook posts regarding vaping have a large proportion of advertisements by vaping companies [11], we removed ads from the CrowdTangle dataset to better identify how the public frames vaping, instead of viewpoints promoted by ads. A total of 91,3726 (74.2%) of the posts were advertisements, resulting in 31,7653 posts that were not ads. More information on how we detected ads is provided below.

   **Advertisement Classifier** Given the prominence of vaping ads on Facebook, we built a vaping advertisement classifier for our CrowdTangle data. We use a BART large model for NLI-based Zero Shot Text Classification [19] trained on the MultiNLI (MNLI) dataset. BART is a denoising autoencoder for pretraining sequence-to-sequence models. BART is trained by corrupting text with an arbitrary noising function, and learning a model to reconstruct the original text. The classifier was used to classify individual Facebook posts in the CrowdTangle data (0=not vaping ad, 1=vaping ad). We extracted a random sample of 2000 posts from the classified data. We then selected three content experts who had published at least ten peer-reviewed articles in the last three years around vaping. Two of these content experts manually classified these 2000 posts (0=not a vaping ad, 1=vaping ad). Within these posts, the distribution was as follows, 0:563, 1:1437, yielding >90% agreement with the machine-assigned labels. Examples of ads: *The Smok Alien Kit is now available in blue! Vaping on Blue Raspberry by Emoji Liquids, Get 20% Off 3 Packs of Mistic's Flavored E Cig Cartridges*.

**Dynamic Topic Modelling** To quantitatively analyze the evolution of vaping-related topics over time and platforms, we used dynamic topic modeling (DTM). While topic modeling analyzes documents to learn meaningful patterns of words, for documents collected in sequence, DTMs capture how these patterns vary over time [2]. DTMs use state space models on the natural parameters of the multinomial distributions that represent the topics [2]. We use BERTopic [7] to visualize the DTMs over time and across platforms.

**Semantic Change** Understanding how words central to our research question, such as *vaping*, change their meanings over time is key to comprehending the evolving framing of vaping. Previous work [8] demonstrated how the word *gay* shifted in meaning from *cheerful* to referring to homosexuality from the 1900s1990s. In a similar way, in this paper we use word embeddings as a diachronic (historical) tool to quantify the semantic change around the word *vaping* over time, thus providing insight into how framing around vaping shifted over time during the period (2009-2021) as well as between media sources. We use an updated version of Gensim word2vec models [1] based on HistWords [2] to provide insight on how *vaping* is framed from 2009-2021, for both CrowdTangle and Media Cloud data. For brevity, we removed words in the fringes of the final output, irrelevant to our RQ, such as *can*, *check*, *need*, and *let*. **Question Answering** Question answering can help us to understand how news and social media *answer* the same questions about vaping, perhaps revealing differences in vaping frames. For example, news media may be more likely to present vaping harms compared to social media. We used BERT [3] fine-tuned on the SQuAD v1.1 dataset [24] for answer extraction. The model was applied separately on news and social media. Questions were developed based on input from content experts, indicated above. Each content expert first developed a list of ten questions separately. The three experts then discussed their lists to result in a final list of four questions that were broadly similar across all three original lists, and final questions are as follows: Can vaping help you quit smoking regular cigarettes? What are the health consequences of vaping? Why are teens vaping? What is the biggest concern with vaping? We highlighted one question at a time and fed it to the model. While we would have preferred to use more than four questions for our question answering analysis, only four questions were agreed upon by the content experts. This is largely due to disagreement among content experts as to what questions should be included, largely resulting from the controversial nature of vaping, and that academics are in disagreement about the harms and merits of vaping [10]. The model extracts answers for the question leveraging on context information in each article or post. To stay within the admitted input size of the model, we clipped the length of the text (title + body text) to 512 tokens. Each question provided one answer per article or post. We randomly sampled 500, 1000, 1500, and 2000 answers per question. We found that a random sample of 1000 answers provided the greatest range and quality of answers, assessed by two reviewers (90% agreement). We thus randomly sampled 1000 answers per question and content experts then selected the top 5, 10 and 20 most representative answers per question, for both news and social media. We found that selecting the top 5 most representative answers provided the least repetition.

---

[1] gist.github.com/zhicongchen
[2] github.com/williamleif/histwords

**Cloze Tests** Following recent work [23], we used BERT [3] and cloze tests to further understand the differences between Media Cloud and CrowdTangle data. Cloze tests represent a fill-in-the-blank task given a sentence with a missing word [27]. For example, *winter* is a likely completion for the missing word in the following cloze task: In the [MASK], it snows a lot. We developed several cloze tests with input from content experts as described earlier. Each content expert first developed a list of ten cloze tests separately. Then the three experts discussed their lists to come up with a final list of four cloze tests: i) The main issue with vaping is [MASK]; ii) The worst thing about vaping is [MASK]; iii) Teens like vaping because it's very [MASK]; iv) Vaping is [MASK] for smoking. We applied BERT on our Media Cloud and CrowdTangle data to identify the differences in the top five results for each cloze test across the media platforms.

## 4   RESULTS

**N-grams** We extracted the frequent n-grams over time to provide an overview of how framing around vaping has changed. Our unigrams and bi-grams did not provide useful results, with the most popular output being *vaping, cigarette, electronic, electronic cigarette, box mod, vape shop*. However, the trigrams provided more fruitful output and we present those results. In the CrowdTangle data, common trigrams in 2009-2018 included *electronic cigarette explodes* and *switch to vape*. These trigrams were likely related to an early case of vape-related injury and transitions from smoking to vaping. After 2018, common trigrams included *food drug administration* and *center disease control*. This shift toward the regulatory aspects of vaping is likely linked to EVALI. Even after EVALI ended in 2020, the narrative around vaping regulation seems to persist in CrowdTangle data. We note the rapid and persistent shift in vaping frames from an occasional news item or smoking cessation tool in 2011-2018; to a regulatory issue in 20192021. In the Media Cloud data, the most common trigrams from 2009-2018 were *food drug administration* and *center disease control*, which are indicative of news media focusing on the regulatory aspects of vaping. In 2019, the most

8         Keyu Chen, Marzy Babaeianjelodar, et alcommon trigrams were *ban flavored e-cigarettes* and *vitamin E acetate*, which likely reflect EVALI. In 2020-2021, there was a renewed focus on regulating vaping, with similar trigrams as in 2009-2018. Overall, we can conclude that while CrowdTangle vaping frames may start around a mode of smoking cessation, framing over time shifted toward news media framing of vaping as a regulatory issue.

**Topic Modelling** Figure 1b represents dynamic topic model results for the Media Cloud data. The five most frequent words for each topic are presented in the Figure legend. Topic proportions are minimal from 2009-2019. From 2019 onward, we observe increases in topics around the COVID-19 pandemic (Topic 7) and EVALI (Topic 1). Topic proportions seem to plateau towards the end of our data. Figure 1a denotes results for CrowdTangle. We noted occasional spikes for topic proportions post-2019, around vaping as a safer alternative to conventional cigarettes (Topic 7), and COVID-19 (Topic 10). Topic proportions appear to plateau toward the conclusion of the analysis period. Broadly, we noted the prevalence of COVID-19-related topics across news and social media [13]. Most importantly, Facebook data contained discussion around vaping as a combustible cigarette alternative, but news media centered instead on vaping harms and regulation.

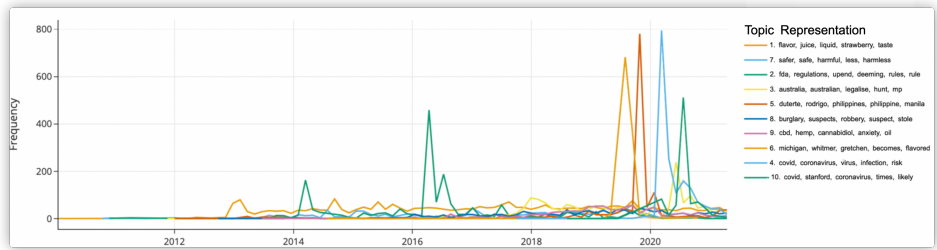

(a)

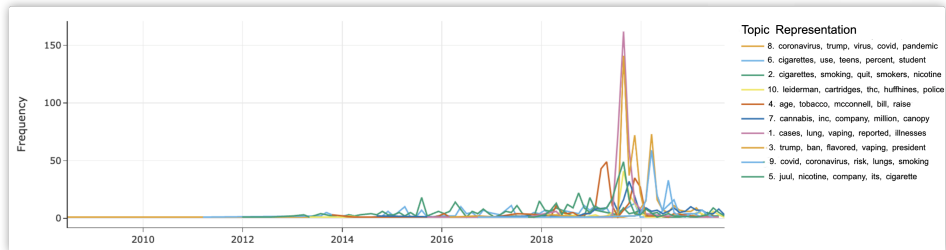

(b)

Fig.1: Dynamic topic model results for CrowdTangle 1b and Media Cloud 1a data from 2009 to 2021

**Semantic Change** Figure 2a and 2b visualizes the semantic change over time for the word *vaping*, from 2009 to 2021. This analysis demonstrates how the meanings of words shift over time, which helps the understanding of semantic changes around vaping. In Figure 2a for the CrowdTangle data, we observe that in 2010, *vaping* was close in meaning to words such as *good* and *weekend*, indicating that vaping was perceived as a fun and enjoyable weekend activity. In 2011, *vaping* moved closer in meaning to *products*, perhaps indicating an increase in the growth and in the availability of vaping products. During the 2017-2021 period, there was a shift toward words such as *teens*, *young*, *epidemic*, *risks*, and *dangers* which indicates a change in meanings towards vaping, toward vaping being dangerous and perhaps responsible for the youth vaping epidemic, likely motivated by EVALI. During the same time period, there was also a shift toward *quit*, *alternative*, and *smoking*, indicating a shift in meanings toward vaping as an alternative to combustible cigarettes and a smoking cessation tool. We note the opposing meanings around vaping that seem to co-occur (cessation tool vs. risky product), which could be indicative of the emergence of polarized public views regarding vaping. Figure 2b, shows the semantic change for Media Cloud data. There were too few vaping-related articles in the 2008-2012 period, and thus Figure 2b shows the semantic change for 2013-2021. We can see that in 2013, *vaping* was close in meaning to *smoke, cities, bar*, perhaps indicative of discussions around indoor vaping bans. In 2015, *vaping* moved closer to *devices* and *regulation*, reflecting the shift in discussion to vaping regulation. From 2016-2021, there was a shift towards *teens* and *youth*, indicative of youth vaping epidemic discussions, and there was also a simultaneous shift towards *juul, flavors, ban*, likely around the planned ban on flavored vaping products. The news media framing of vaping has shifted over time in line with emergent regulatory trends; such as; flavored vaping bans, with little discussion around vaping as a smoking cessation tool. Social media discussions were far more varied, with transitions toward vaping as both a public health risk and a smoking cessation tool.

| Media Cloud (probability) | CrowdTangle (probability) |
|---|---|
| bad (0.148) | better (0.582) |
| safe (0.105) | bad (0.036) |
| not (0.093) | not (0.016) |
| dangerous (0.067) | substitute (0.014) |

Table 1: The top four candidate words ranked by BERT probability for the cloze test "*Vaping is [MASK] for smoking*" for Media Cloud and CrowdTangle data.

| Media Cloud | CrowdTangle |
|---|---|
| effective interventions remain elusive | if you can bypass the liver |
| Vaping can help people quit smoking | consumers avoid all vaping products |
| giving me life | damage their lungs |
| If you cannot quit | don't vape |
| Vaping helps smokers quit | no vaping |

Table 2: Question-answering results for *Can vaping help you quit smoking regular cigarettes?* for Media Cloud and CrowdTangle.

**Question Answering** We present answers to *Can vaping help you quit smoking regular cigarettes?* and the top five most representative answers across news and social media in Table 2. We only found useful results for the indicated question. Other questions did not provide useful results and thus are not presented. For example, the question, *Why are teens vaping?* provided answers such as *bullying* and *health scare.* There was a clear difference in vaping frames

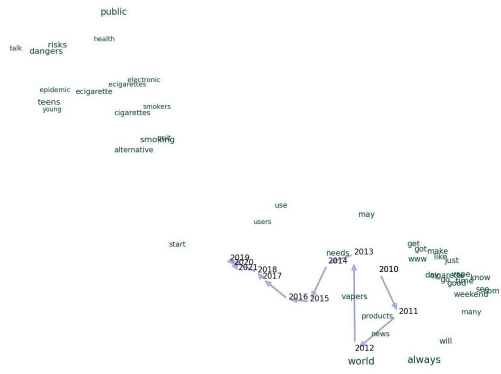

(a) Semantic change for *vaping*, on the CrowdTangle data

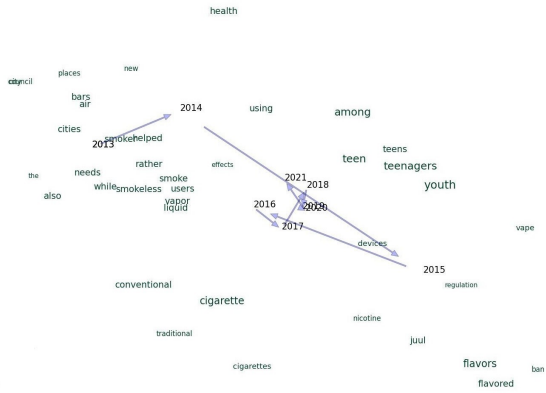

(b) Semantic change for *vaping*, on the Media Cloud data

Fig.2: Two-dimensional visualizations of the semantic change for the word *vaping*, from 2010 to 2021, for CrowdTangle and for Media Cloud data.

for Media Cloud vs CrowdTangle. News media tends to view vaping as a danger, with responses such as *damage their lungs* and *consumers avoid all vaping products*. Social media displays more balanced answers, such as, *effective interventions remain elusive*, *vaping can help people quit smoking*, and *Vaping helps smokers quit*. News media tends to portray vaping as a health harm while social media includes discussion around vaping as a smoking cessation tool.

**Cloze Tests** We use cloze tests to gauge the aggregate framing around vaping, across media platforms. Table 2 shows the cloze test results for the probe: "Vaping is [MASK] for smoking". We can see that news media views vaping as *bad* or *dangerous*, but also *safe*. In CrowdTangle, we note a range of vaping frames of vaping, with vaping being a *substitute* and *better* for smoking. While both news and social media frame vaping as both dangerous and safe, reflecting diverse viewpoints, only social media characterizes vaping as an alternative to cigarettes. We speculate that news media are less likely to report on vaping as a smoking cessation tool, unlike social media, which may promote that view along with a range of other views. Other cloze tests did not provide useful information. For example, for the probe: "The main issue with vaping is [MASK]", we found results such as *the* and *va* across the media platforms, which are not useful.

## 5   DISCUSSION

**Implications of our findings** Our RQ was to explore the broad differences between news media and social media regarding vaping. A strength of our work is how the different techniques we have applied validate each other. For example, the semantic change and n-gram results detail similar shifts from vaping as a smoking cessation tool to a regulatory issue. Results suggest that social media tracks the evolution of vaping as a social practice, while news media reflects more risk based concerns driven by evolving regulation and public health narratives. Vaping seems to have evolved away from mimicking smoking to being a related but separate practice with its own rituals and subculture so to some extent we can see how social media discussions reflect that evolution, particularly with regard to vaping as a product category. As the vaping landscape continues to evolve, it is possible that more vaping regulation inimical to smoking cessation is proposed. We suggest vaping regulation that clarifies the role of vapes as a smoking cessation tool. To improve framing around vaping, minimize youth vape use, and possibly facilitate improve smoking cessation through vaping, policy makers can conduct tailored interventions and can start communication campaigns to counter possibly negative media rhetoric around vaping [5]. An example intervention can use brief exposure to evidence-based information about vapes, perhaps reducing false beliefs about vaping. Similar interventions can use observed differences in news and social media to improve evidence-based interventions in various platforms.

**Limitations** Our findings relied on the validity of data collected with our search terms. We used Media Cloud and CrowdTangle to search for all articles relevant to vaping, and our data contained text aligned with how vaping is framed. We are thus confident in the comprehensiveness of our data. We note that our data sources, Media Cloud and CrowdTangle are not representative enough to be considered public discourse. Thus, we indicate that our data may not be generalizable to how vaping is framed in the US. We were not able to obtain statistics about how many times an article was read or shared, or to control for news outlets that are more widely read compared to smaller regional news outlets. We were also not able to distinguish between bias-free publications and opinion/commentary articles. Findings may also not apply to other related issues that are also heavily politicized (e.g., abortion) or other contexts (e.g., vaping frames in Europe). We also note the limitations of BERT [28], such as its inability to learn in few-shot settings. Such model limitations hampered our ability to analyze how subgroups of vapers are framed, such as LGBT+ vapers, around which there were relatively few news articles.

**Conclusion** We found that news media frames of vaping transitioned over time per emergent regulatory trends, such as; flavored vaping bans; with limited discussion around vaping as a smoking cessation tool. Framing of vaping in social media vaping was more varied; with transitions toward vaping both as a public health harm and a smoking cessation tool. Stakeholders may utilize our findings to intervene around vaping where necessary; and to design communications campaigns, among other measures, possibly aiding smoking cessation, and reducing youth vaping.